\documentclass[a4paper,11pt]{article}
\usepackage{jinstpub} % for details on the use of the package, please see the JINST-author-manual
\usepackage{lineno}
\usepackage[thinspace,thickqspace,amssymb,binary,italian]{SIunits}

\title{\boldmath A prototype electromagnetic calorimeter for the MUonE experiment: status and first performance results}

\author{E. Spedicato}
\collaboration[c]{on behalf of the MUonE collaboration}
\affiliation{INFN, Sezione di Bologna,\\
Viale Berti-Pichat 6/2, Italy}
\affiliation{University of Bologna,\\
Via Irnerio 46, Italy}

\emailAdd{espedica@bo.infn.it}

\abstract{The MUonE experiment proposes a novel approach to determine the leading hadronic contribution to the muon g-2, from a precise measurement of the differential cross section of the $\mu-e$ elastic scattering, achievable by using the CERN SPS muon beam onto atomic electrons of a light target. The detector layout is modular, consisting of an array of identical tracking stations, each one made of a light target and silicon strip planes, followed by an electromagnetic calorimeter made of $PbWO_4$ crystals with APD readout, placed after the last station, and a muon filter. The scattering particles are tracked without any magnetic field, and the event kinematics can be defined in a large phase space region from the expected correlation of the outgoing particle angles. The ambiguity affecting a specific region, with electron and muon outgoing with similar deflection angles, can be solved by identifying the electron track as the one with extrapolation matching the calorimeter cluster or the muon track by associating it to hits in the muon filter. The role of the calorimeter will be important for background estimate and reduction, and to assess systematic errors, providing some useful redundancy and allowing for alternative selections.
Beam tests are carried out at CERN with a prototype calorimeter to determine its calibration with both high energy ($\unita{20-150}{\giga\electronvolt}$) and low energy electrons ($\unita{1-10}{\giga\electronvolt}$). In late summer a pilot run is scheduled with up to three tracking stations and the calorimeter integrated within a common triggerless readout system. The main motivations for the MUonE calorimeter are discussed, and the status and first performance results will be presented.}


\begin{document}
\maketitle
\flushbottom

\section{Introduction}
\label{sec:intro}

Nowadays, one of the most puzzling question in particles physics concerns the muon magnetic moment anomaly $a_{\mu}=g-2/2$. In these last 20 years, several efforts have been put in the study of this quantity. In 2006, the measurement of the E821 experiment~\cite{ref:bnl} at BNL resulted to deviate $3.7\sigma$ from the reference theoretical estimate. After fifteen years, in 2021~\cite{ref:g20} and again in 2023~\cite{ref:g2}, the $g-2$ experiment at Fermilab confirmed this result, bringing the discrepancy first to $4.2\sigma$ and then to $5.0\sigma$. However, the theoretical landscape changed in the last five years.\\
The standard model contributions to $a_\mu$ come from QED, electroweak and hadronic theory, such that 
\begin{equation}
a_\mu=a_\mu^{QED}+a_\mu^{EW}+a_\mu^{had}.
\end{equation}
While the first two can be precisely evaluated in perturbation theory, the latter cannot. The dominant uncertainty of the prediction is given by this term, in particular by the hadronic vacuum polarization contribution at leading order $a^{HVP}_\mu$. As it cannot be calculated perturbatively, the reference value reported in the 2020 White Paper~\cite{ref:wp} was evaluated through a data-driven method based on the measurement of the cross section of $e^+e^-$ annihilation to hadrons. Until some years ago, it was the only method with an uncertainty comparable with the experiment, but, in 2020, the BMW Collaboration~\cite{ref:lqcd} published a new result for $a^{HVP}_\mu$ based on Lattice QCD, which had for the first time a competitive precision. This resulted to weaken the $a_\mu^{th}-a_\mu^{exp}$ discrepancy, bringing it to $1.5\sigma$. In 2023, the new measurement of the  $e^+e^-\rightarrow had$ cross section from the CMD-3 experiment~\cite{ref:cmd} seems to displace the expected anomaly towards the experimental measurement, in agreement with the BMW result, contributing to get the puzzle more complex.\\
Considered this intriguing situation, it is of paramount importance to improve the theory calculation and clarify the status of the available estimates. The MUonE proposal aims to provide an independent and innovative method to evaluate $a^{HVP}_\mu$ \cite{ref:pr, ref:mue}, based on the master equation \cite{ref:me}:
\begin{equation}
a^{HVP}=\frac{\alpha}{\pi}\int_{0}^{1} dx(1-x)\Delta\alpha_{had}[t(x)],
\end{equation}
which needs as input the hadronic contribution to the running of the electromagnetic coupling in the space-like region of momenta $\Delta\alpha_{had}(t)$, where $t(x)$ is the space-like (negative) squared four-momentum transfer. The running can be determined from the precise measurement of the shape of the differential cross section of the $\mu-e$ elastic scattering on a light target. The CERN M2 muon beam line with $E_{\mu}=\unita{160}{\giga\electronvolt}$ will be used for the purpose. The main difficulty of the method is to reach the required precision in order to be competitive: $\le 0.5\%\,a^{HVP}_\mu$. The challenge will be to keep the systematic uncertainties under control. 
The project has been submitted to the CERN SPS Committee with a Letter of Intent in 2019 \cite{ref:LOI} and the next step will be to write a technical proposal in 2024.

\section{MUonE analysis technique}
The $\Delta\alpha_{had}(t)$ contribution is most easily displayed considering the ratio $R_{had}$ of a cross section including the full running of $\alpha$ and the same cross section with no hadronic running:
\begin{equation}
 R_{had}=\frac{d\sigma(\Delta\alpha_{had}\ne0)}{d\sigma(\Delta\alpha_{had}=0)}
\end{equation}
The four-momentum transfer $t$ of the $\mu-e$ interaction can be directly determined by the electron and muon scattering angles through defined kinematic relations. Thus, the measurement of the shape of the differential cross section in this experiment is based on the precise measurement of the leptons scattering angles. Once it is evaluated, a template fit method has been chosen to extract $ R_{had}$ from data, where $\Delta\alpha_{had}(t)$ is fitted by a two parameters function \cite{ref:GA}:
\begin{equation}
 \Delta\alpha_{had}(t)=k \left\{ -\frac{5}{9}-\frac{4M}{3t}+\left(\frac{4M^2}{3t^2}+\frac{M}{3t}-\frac{1}{6}\right)\frac{2}{\sqrt{1-\frac{4M}{t}}}\left|\frac{1-\sqrt{1-\frac{4M}{t}}}{1+\sqrt{1-\frac{4M}{t}}}\right| \right\}.
\end{equation}

\begin{figure}[htbp]
\centering
\includegraphics[width=.65\textwidth]{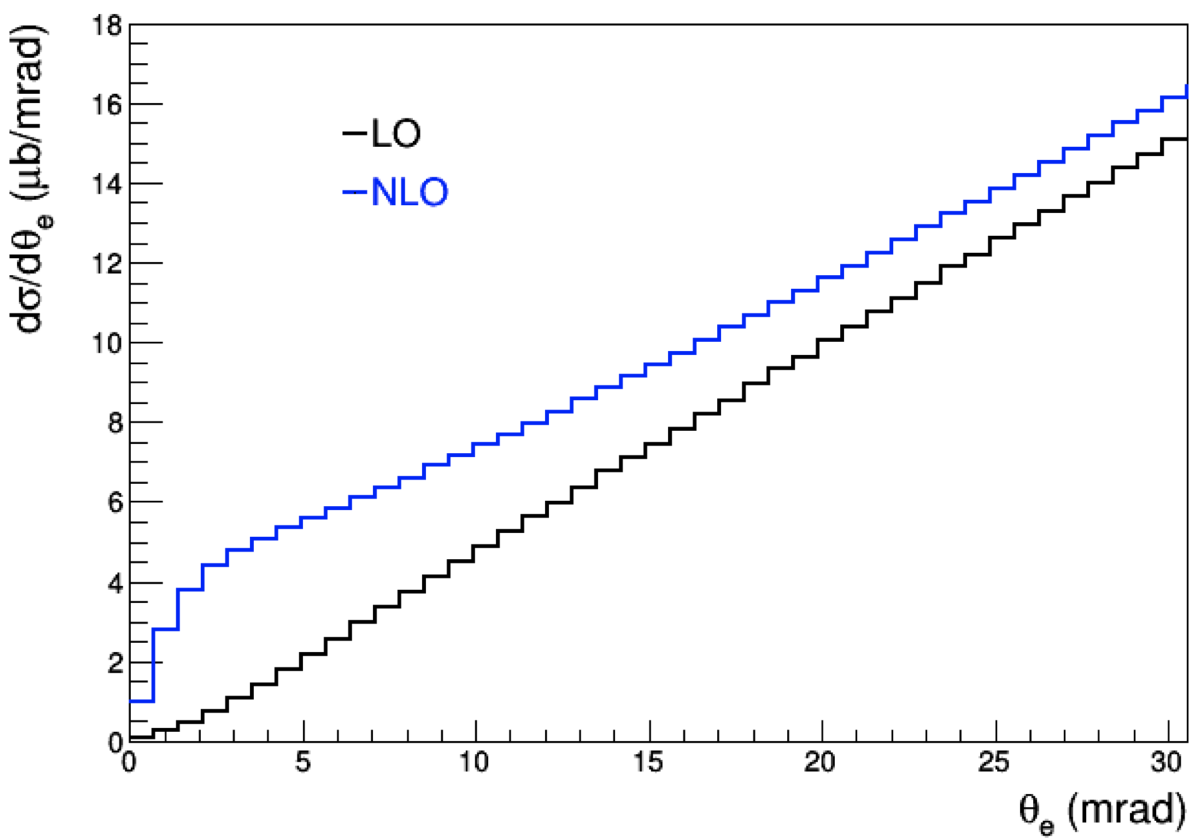}
\caption{Differential cross section of elastic scattering at LO (black) and NLO (blue), as a function of the electron scattering angle.\label{el}}
\end{figure}

This method consists in generating a grid of points $(k,M)$ in the parameters space covering a region of $\pm 5\sigma$ around the expected values, where $\sigma$ is the expected uncertainty. For each pair of values, a template for $R_{had}$ is obtained with the Monte Carlo generator, which then is compared with data/pseudodata calculating:
\begin{equation}
    \chi^2(K,M)=\sum_i \frac{R_i^{data}-R_i^{templ}(K,M)}{\sigma_i^{data}}
\end{equation}
where $K=\frac{k}{M}$, and the minimum $\chi^2$ is found by parabolic interpolation across the grid points. The fit can be done on the distribution of the muon or the electron scattering angle, as well as on their two-dimensional distribution, which gives the most accurate result.\\
The elastic scattering implies a well defined correlation between the two leptons angles and this can be used to help the selection of a clean sample of elastic events. This is fundamental for the precision that is needed. Many effects can distort the signal differential cross section. In Fig.~\ref{el}, the presence of radiative events with real photon emission, expected as a NLO effect, is represented and this modifies the LO elastic signal. Therefore, it is important to find a way to recognize and be able to discard these events.\\

\section{Experimental apparatus}
\begin{figure}[htbp]
\begin{center}
 \includegraphics[scale=0.80]{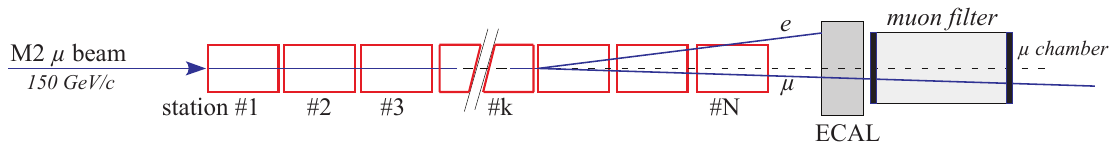}
 \centering
 \caption{Layout of the MUonE experimental setup~\cite{ref:LOI}. \label{app}}
  \end{center}
\end{figure}

The full experimental apparatus is segmented in 40 tracking stations, followed by an electromagnetic calorimeter (ECAL) and a muon filter (Fig. \ref{app}).
Each station behaves as an independent unit and is composed by a $\unita{1.5}{\centi\meter}$ target (Beryllium or Graphite) and 6 silicon strip modules (CMS 2S modules \cite{ref:cms}). With this configuration, in three years of data taking it is possible to reach the target statistical sensitivity with an integrated luminosity of $\unita{1.5\times10^{7}}{\nano\barn}^{-1}$. The challenge will be to keep at the same level the systematics, as multiple scattering, knowledge of the average beam energy to few $\unita{}{\mega\electronvolt}$, alignment and intrinsic angular resolution.\\
The electromagnetic calorimeter (ECAL) prototype is composed by 25 cells of $PbWO_4$, with a total surface of $\unita{14\times 14}{\centi\meter}^2$. Signals arrive to avalanche photo-diodes (APD) which are read out by two Front-end boards connected to an FC7 board. To calibrate and control APDs signals, a laser pulse system (at $\unita{450}{\nano\meter}$) is provided.\\

\section{The electromagnetic calorimeter}
\begin{figure}[htbp]
\centering
\includegraphics[width=.4\textwidth]{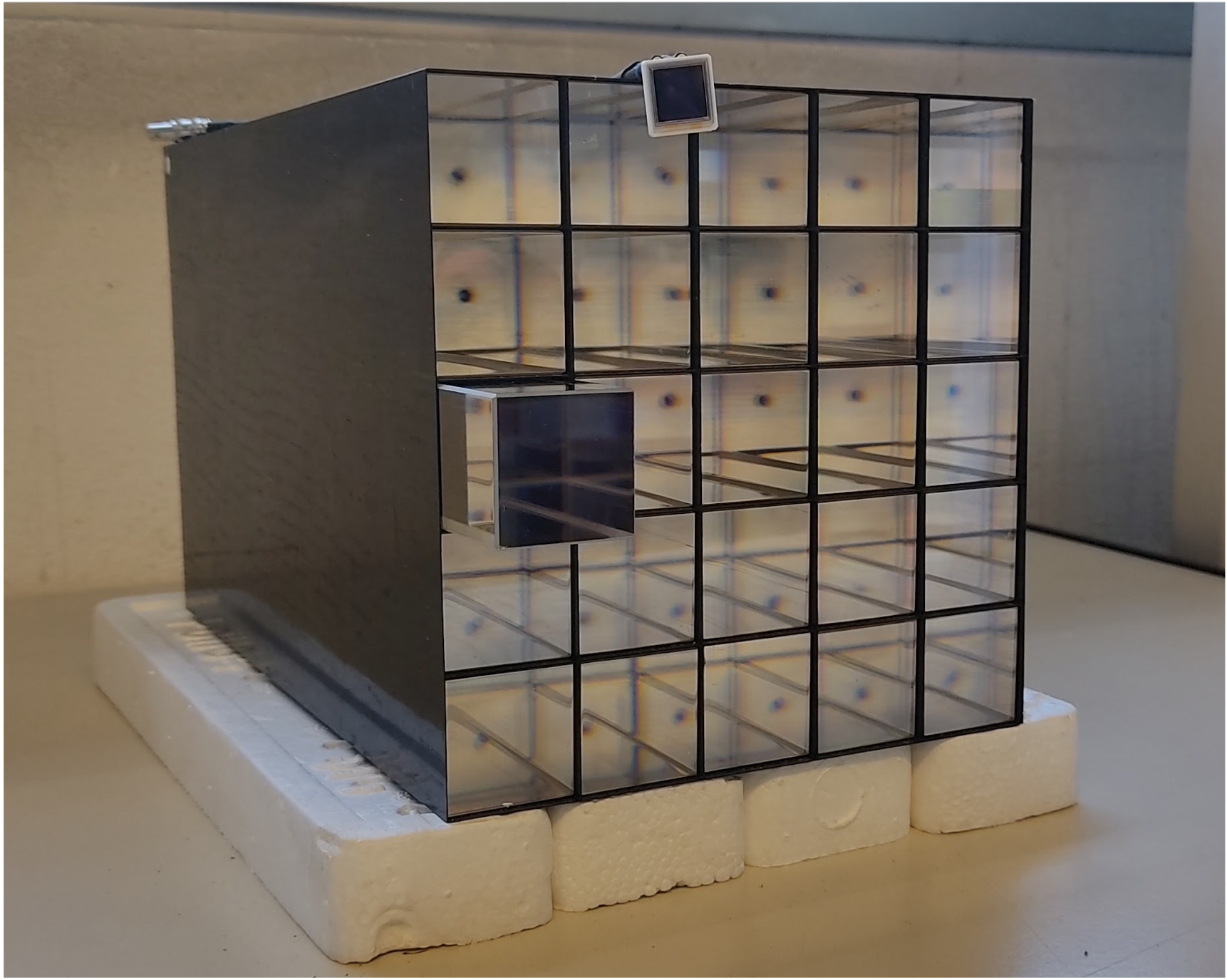}
\qquad
\includegraphics[width=.4\textwidth]{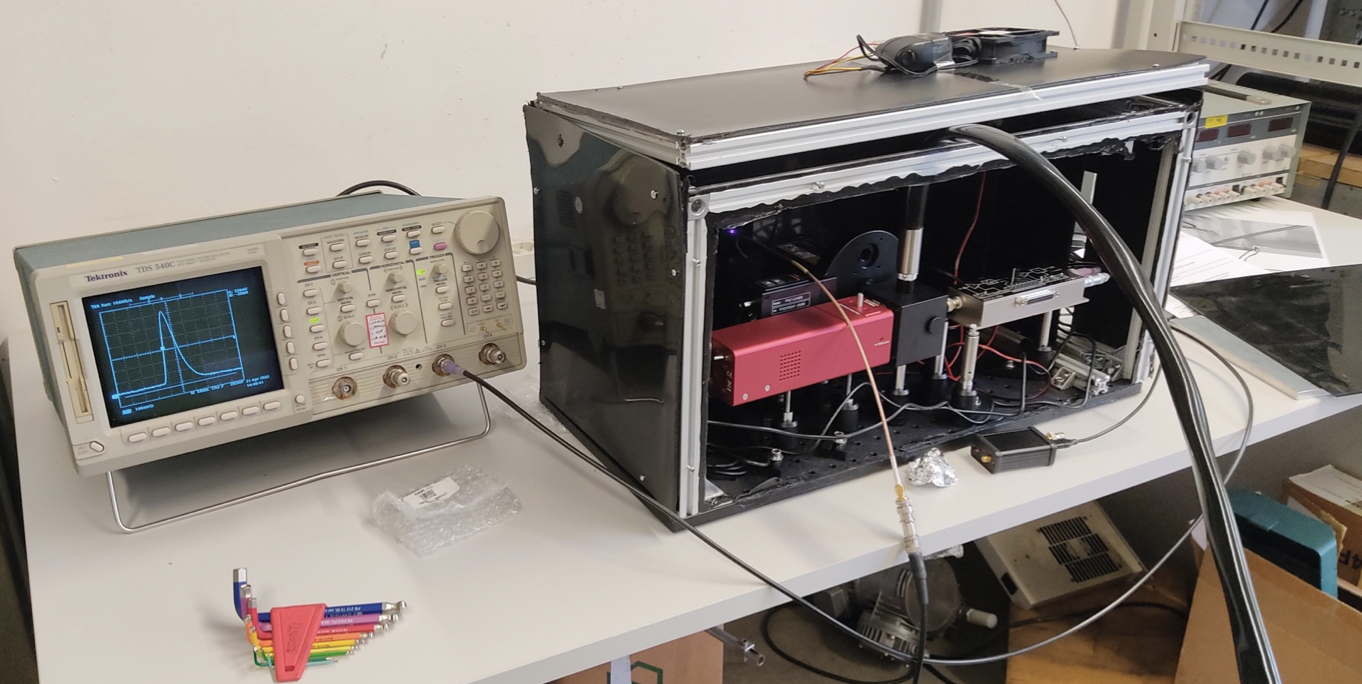}
\caption{(Left) The prototype calorimeter, with the 25 $PbWO_4$ crystals and the APD (shown on top) for the readout; (Right) the $\unita{450}{\nano\meter}$ laser system for the APDs calibration.\label{fig:calolaser}}
\end{figure}

\begin{figure}[htbp]
\centering
\includegraphics[width=.55\textwidth]{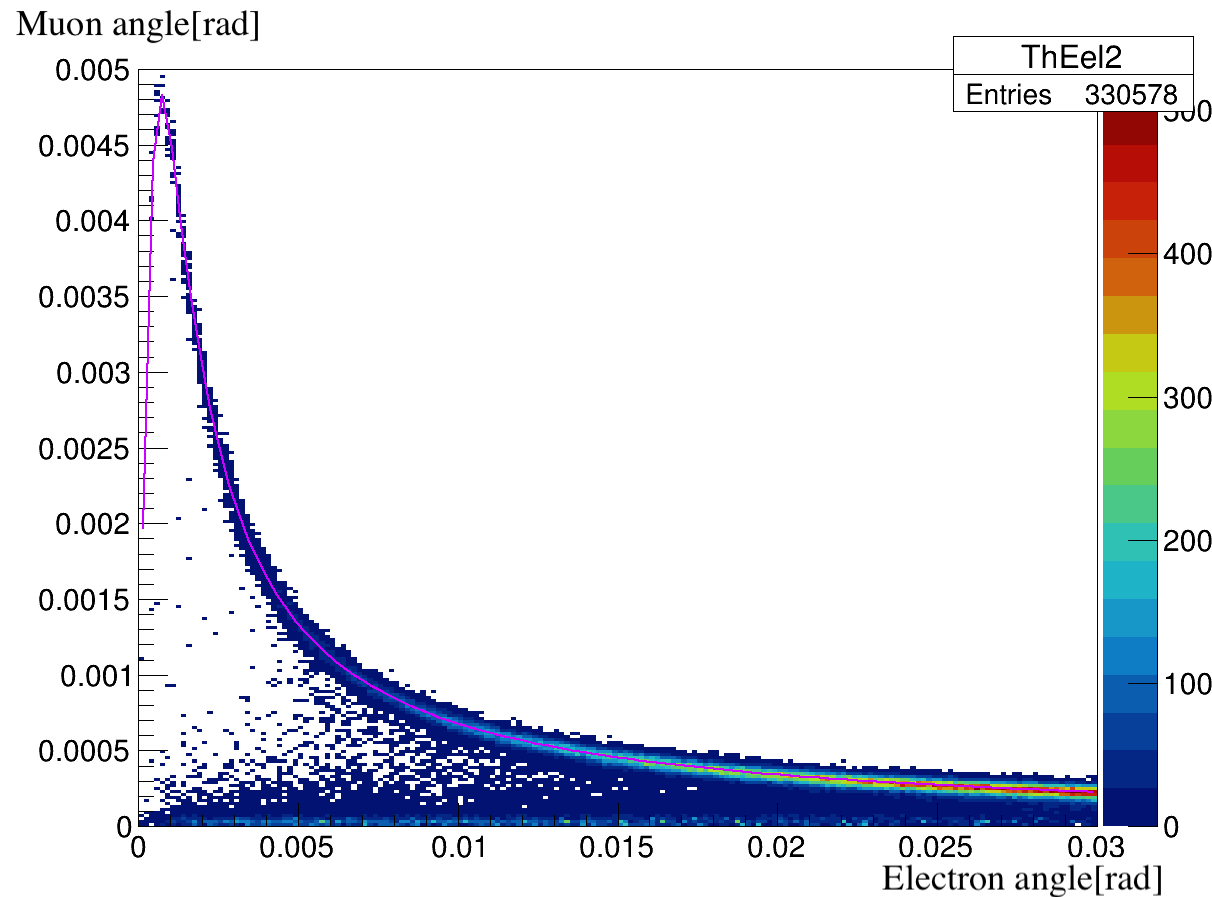}
\qquad
\includegraphics[width=.52\textwidth]{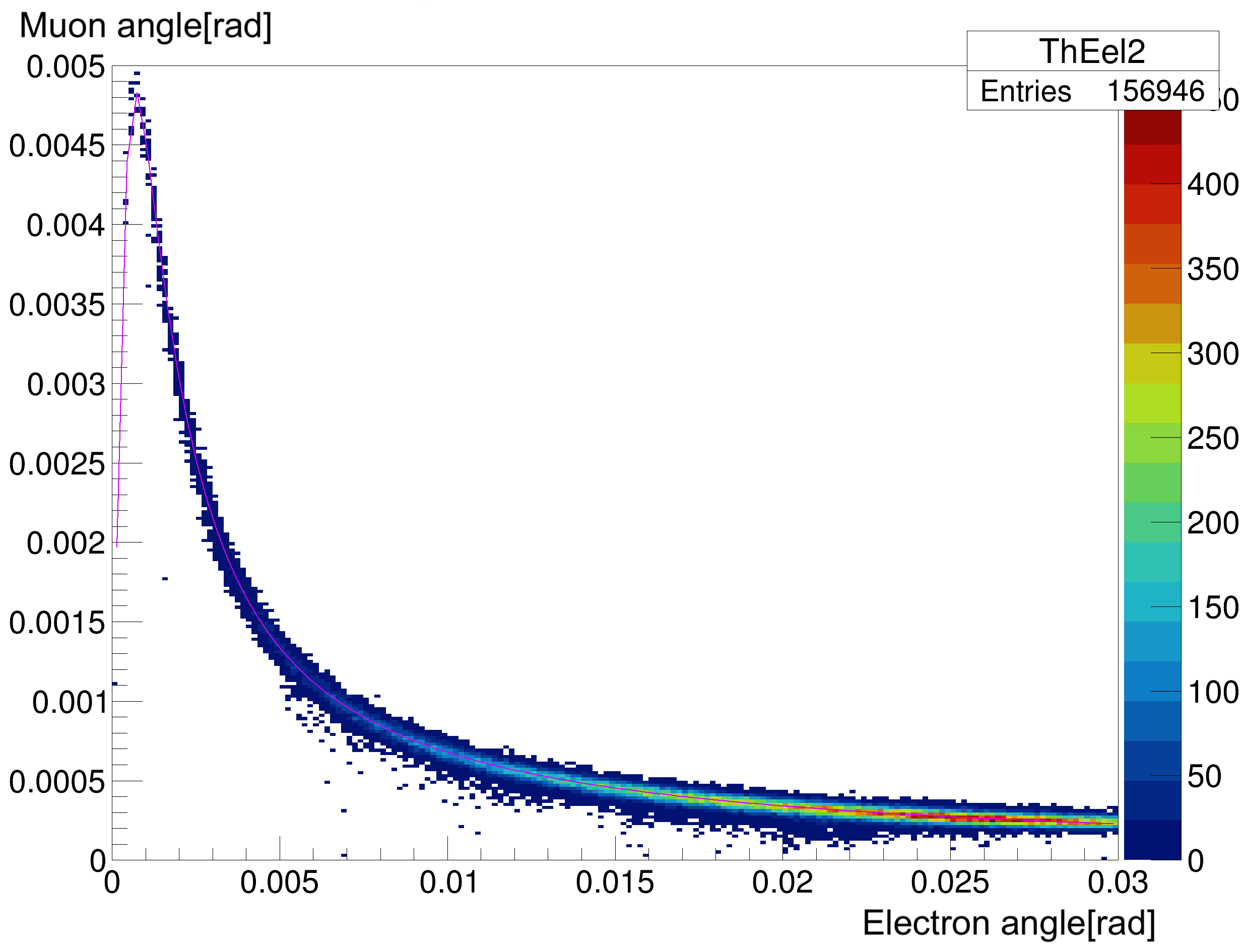}
\caption{Correlation between the muon and electron scattering angles without any cuts (top) and with the ECAL based cuts (bottom) from fast MC simulation of the MUonE Test Run setup. The superimposed violet line represents the expected correlation between the muon and electron scattering angles in perfectly elastic events \cite{ref:tesi}.\label{fig:cuts}}
\end{figure}

\begin{figure}[htbp]
\centering
\includegraphics[width=1.00\textwidth]{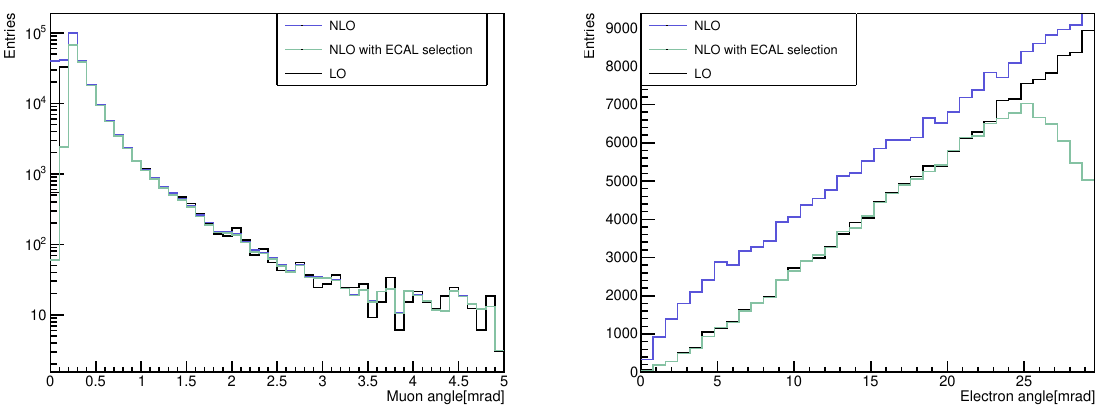}
\caption{Distributions of muon (left) and electron (right) scattering angle from simulated events from the MESMER MC, with fast simulation of the MUonE Test Run setup. NLO distributions are shown before the calorimetric selection (blue histogram) and after it (green histogram). The black histograms show the LO prediction \cite{ref:tesi}.\label{ref:siena}}
\end{figure}

The prototype of the electromagnetic calorimeter is shown in Fig. \ref{fig:calolaser}, together with the laser system that is used to calibrate the APDs.
The prototype is based on the CMS calorimeter, in particular $PbWO_4$ crystals correspond to the crystals used in the CMS endcap calorimeter. However there is a difference in the size of the APDs, that for MUonE are larger ($\unita{10\times10}{\milli\meter}^2$). Therefore, it is expected that the observed signal size and the energy resolution will be affected.\\
According to the hardware specifics, the signal level expected for a $\unita{150}{\giga\electronvolt}$ electron (the highest energy in the MUonE experiment) is
\begin{equation}
    \unita{150,000}{\mega\electronvolt}\times 9\frac{photoelectrons}{\unita{}{\mega\electronvolt}} \times 50 = 6.75\times10^7 \sim 28723\,ADC\,counts
    \label{eq}
\end{equation}
where $50$ is the gain. The system noise is dominated by the CMS multiple-gain preamplifier (MGPA) \cite{ref:mgpa}, a three-gain channel CMOS chip, and its knowledge is important to determine the ECAL capability of identifying the $\unita{160}{\giga\electronvolt}$ beam muons, that release on average $\sim\unita{700}{\mega\electronvolt}$. The expected energy resolution for those particles is $\sim 5-7\%$.\\
The role of the ECAL in MUonE is mostly to use it as a control system for the tracker, helping in assessing the systematics. Some examples of its usage are listed below:
\begin{enumerate}
    \item Check the elastic sample selected by the tracker, in particular in the last stations, where the majority of the particles scatter in ECAL acceptance;
    \item Help to describe and recognize the main background (pair-production) and radiative events, which break the perfect correlation between the muon and electron angles;
    \item Help in identifying the scattered electron, associating an ECAL energy cluster to a matched track, resolving the kinematical ambiguity for small scattering angles.
\end{enumerate}
The possible application of the ECAL to the elastic event selection has been studied in the context of the Test Run setup, initially foreseen to be conducted in 2021 \cite{ref:tesi}. A fast simulation has been developed based on the GFLASH  parametrization used in CMS \cite{ref:gflash,gflash2}. It has been used to study the behaviour of the ECAL and its capability of selecting a clean sample of elastic events. The cuts applied on a $\mu-e$ scattering sample generated by a MESMER MC generator at NLO \cite{ref:mc1,ref:mc2}, therefore including radiative events, are all ECAL based \cite{ref:tesi}. The most useful calorimeter information which have been used are:
\begin{enumerate}
\item The reconstructed energy in the $3\times3$ cluster of cells around the impacted one: $E_{3\times3}$ (only nine such clusters can be built, as 16 crystals lie in the detector borders);
\item The centroid of the electromagnetic shower, which estimates the impact position of the showering particle: $\vec{r_C}$.
\end{enumerate}
The first cut consists in rejecting the events with $E_{3\times3}<\unita{1}{\giga\electronvolt}$, which mainly reduce the number of events with a large electron scattering angle, not interesting for the final analysis as strongly affected by experimental perturbations and not carrying sensitive information on the running of $\alpha$. Among the rejected ones, there are also radiative events with electrons that, after emitting soft photons, scatter at an angle smaller than the predicted elastic one and populate the low $\theta_{\mu}$ region. The second cut is based on the mean energy fraction expected for the scattered electron in elastic events $E_{3\times3}/E^{th}_{el}$, where $E^{th}_{el}$ is the energy predicted theoretically at the corresponding scattering angle $\theta_{el}$, measured event-by-event by the tracking station \cite{ref:mue}. The fraction of energy released by an electron in a $3\times3$ array of crystals is $\sim 95\%$ of the overall true energy. If the true energy of the event does not correspond, within some limits, to the predicted elastic one $E^{th}_{el}$, the fraction $E_{3\times3}/E^{th}_{el}$ will deviate from the expected fraction released and the event is likely to be radiative. The third cut is based on the shower centroid. It can be compared to the extrapolated track position at the ECAL front surface. Cutting on the centroid position is effective when radiative events are present and the photon energy is high enough to shift significantly the shower centroid from the electron impact point.\\
Fig.\ref{fig:cuts} shows the correlation plot of the lepton scattering angles from the NLO MC sample before the ECAL selection (on top) and after it (on bottom), with the violet line representing the theoretical elastic curve.
Looking at the scattering angle distributions of the two leptons before and after the ECAL selection (Fig.\ref{ref:siena}), it is evident that the application of the selection criteria moves the event distributions towards the LO ones, in particular the electron angle. The muon scattering angle instead is a robust observable even with radiative events; as a matter of fact the NLO distribution after the selection cuts mostly overlaps with the LO one.

\section{Test runs}
\begin{figure}[htbp]
\centering
\includegraphics[width=.45\textwidth]{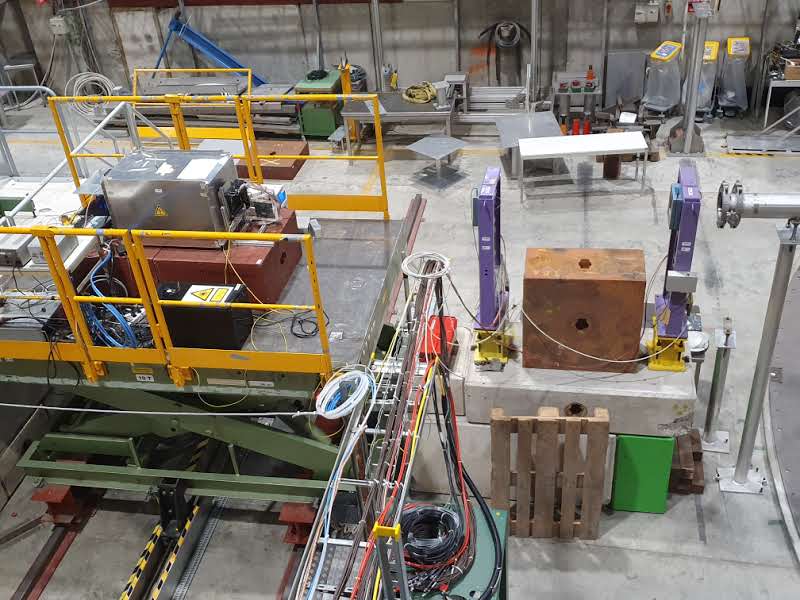}
\qquad
\includegraphics[width=.45\textwidth]{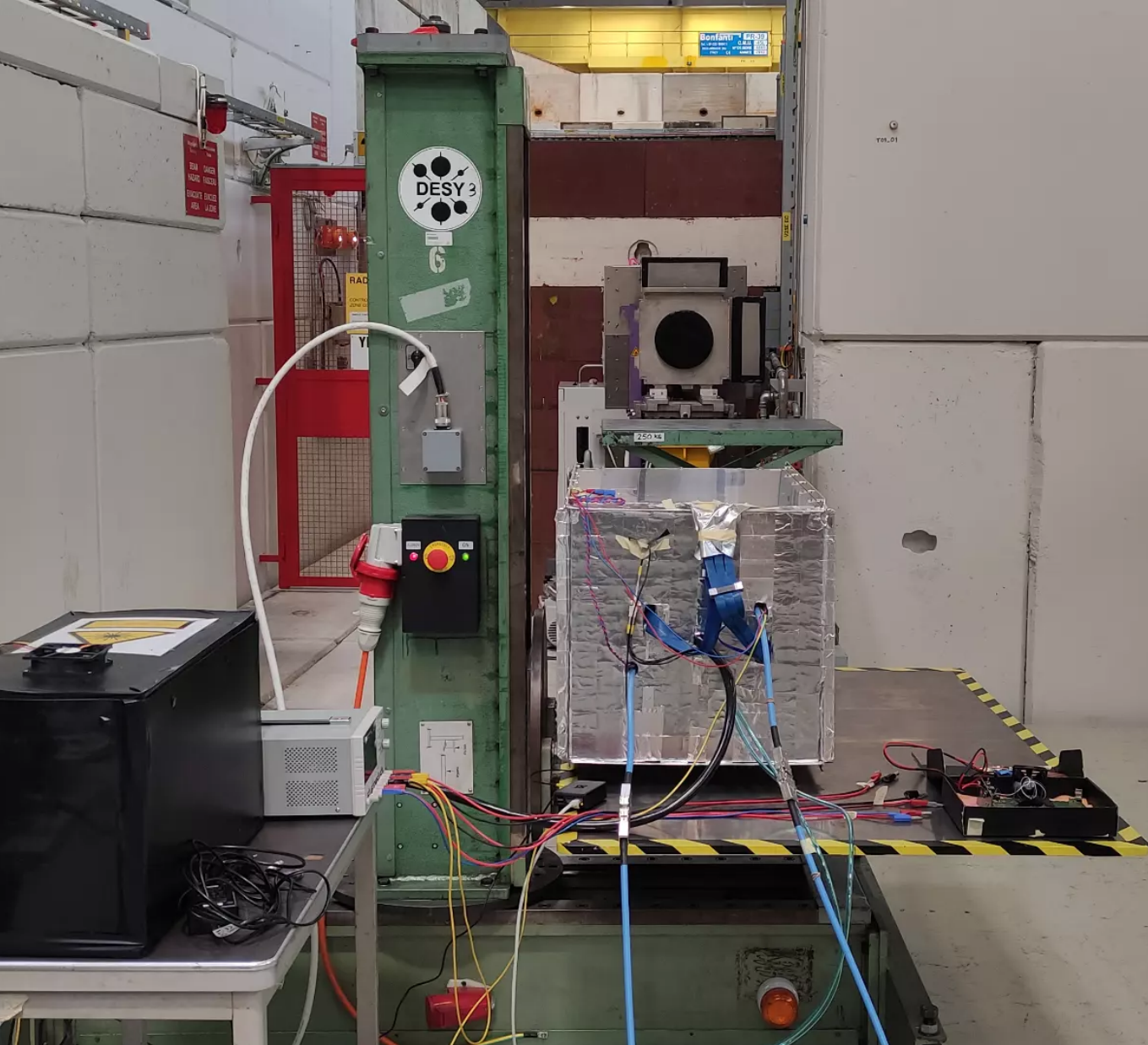}
\caption{H2 area on the left and T9 area on the right where the calibration tests on the ECAL were carried out.\label{t9h2}}
\end{figure}

\begin{figure}[htbp]
\centering
\includegraphics[width=0.7\textwidth]{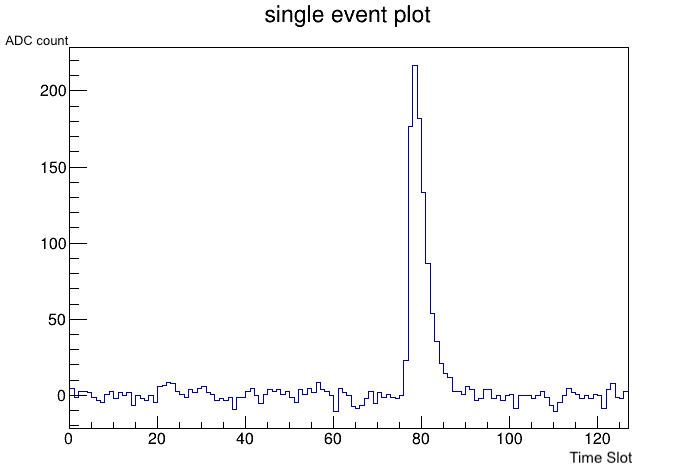}
\caption{Time evolution of the ADC signal in a single crystal for $E_e=\unita{1}{\giga\electronvolt}$\label{shape}.}
\end{figure}

\begin{figure}[htbp]
\centering
\includegraphics[width=1.\textwidth]
{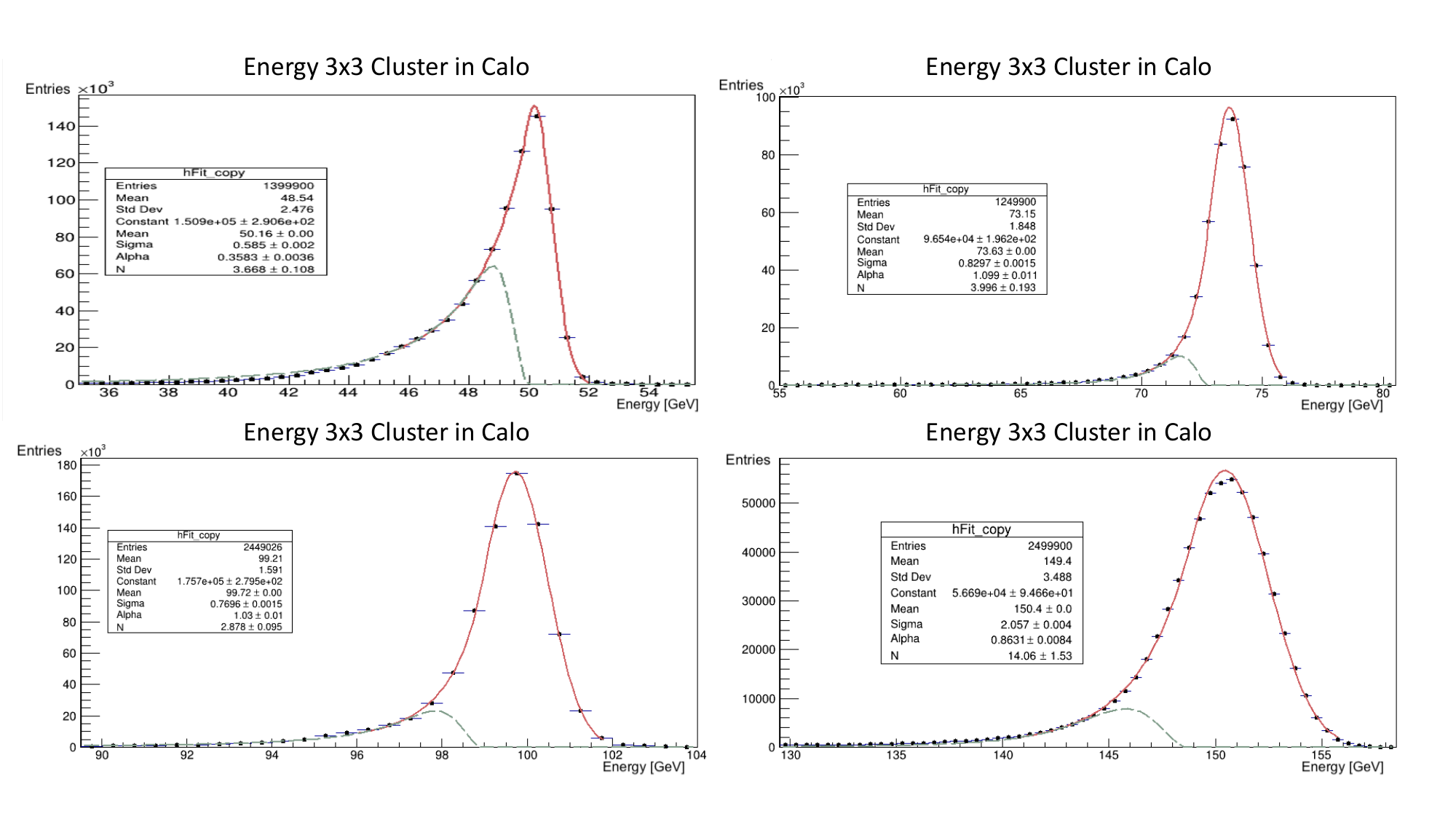}
\caption{Cluster energy distributions, fitted with a Crystal Ball function in red. The green line represents the Crystal Ball tail, where the Gaussian peak was subtracted. Top left: $E_e=\unita{50}{\giga\electronvolt}$; top right $E_e=\unita{75}{\giga\electronvolt}$; bottom left $E_e=\unita{100}{\giga\electronvolt}$; bottom right $E_e=\unita{150}.{\giga\electronvolt}$. \label{dist}}
\end{figure}

\begin{figure}[htbp]
\centering
\includegraphics[width=0.80\textwidth]{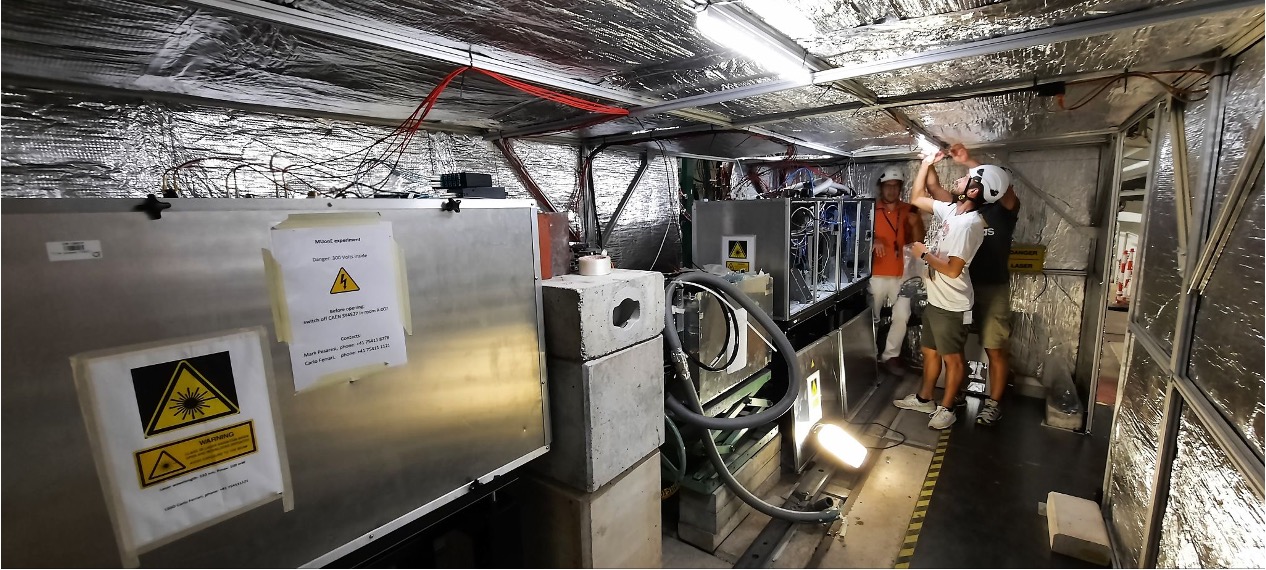}
\caption{Picture of the isolating structure containing the tracking stations and the ECAL during the August-September pilot run.\label{sett}}
\end{figure}

Several test runs have been carried out between 2017 and 2023, to check the feasibility of the MUonE experiment and the behavior of the detectors.\\
In 2017 a first test was performed to study multiple scattering \cite{ref:mcs}. In 2018 a second one was set up at the CERN M2 beam line in order to study the capability of selecting elastic events \cite{ref:2018}. The used detectors had a worse resolution and working conditions with respect to the expected final ones, however we succeeded in selecting a first clear sample of elastic events.\\
During summer 2022, a first test on the ECAL was done at the CERN T9 beam line to verify the response of the detector to low energy electrons. More complete tests have been carried out in 2023 (e.g. application of optical grease between APDs and crystals). The main goal was to calibrate the ECAL with different ranges of electrons energy:
\begin{itemize}
    \item $\unita{20-150}{\giga\electronvolt}$ electrons at the H2 CERN beam line (Fig. \ref{t9h2} left);
    \item $\unita{1-10}{\giga\electronvolt}$ electrons at the T9 CERN beam line (Fig. \ref{t9h2} right).
\end{itemize}
The APD voltages were finely tuned to achieve the best equalization of the signal within $10\%$ in all of them. During data taking, the calibration was done outside the beam spill, as each crystal was irradiated with the light coming from the laser. The output of each APD is sampled every $\unita{25}{\nano\second}$ and that corresponds to one time slot (TS). For a single ECAL event, 128 TS are collected per crystal. The calorimeter operates in a self-trigger mode, saving just the events with the sum of the ADC counts in the 25 crystals exceeding a fixed threshold. The shape of the signal in time is shown in Fig.\ref{shape} for a $\unita{1}{\giga\electronvolt}$ electron. The evolution has a quick rise (in about two time slots) and then a slower decay, with a delay between the start of data acquisition and the signal of about $70$ TS for low energy electrons. The first $30$ TS are used to evaluate the pedestal subtraction and the noise.
During these tests, the ECAL was aligned along the direction of the beam, with a mobile platform that allowed the displacement of the detector along the horizontal and the vertical directions. At each energy, data for calibration were taken with the beam centered in each crystal. To evaluate the total charge collected by the APD, the ADC signal is fitted in a limited region around the maximum, with a model introduced by the CMS collaboration \cite{ref:cms_sig1,ref:cms_sig2}. It is a power function multiplied by a dumping exponential:
\begin{equation}
    A(\Delta t)=S(1+\frac{\Delta t}{\alpha\beta})^\alpha e^{-\Delta t/\beta}
\end{equation}
where $S$ is the ADC signal, $\Delta t=t-\bar{t}$ is the difference between a time $t$ and the mean time $\bar{t}$ of the signal peak in unit of TS and $\alpha,\,\beta$ are the shape parameters. The fit is limited in the range $-2\le\Delta t\le 2$. After calibration, the resulting cluster energy distributions at different beam energies are shown in Fig.\ref{dist}.
These plots are obtained selecting events where the sum of the ADC counts collected in all the crystals exceeds a certain threshold. Then, the crystal with the maximum energy is identified and the signal is considered only if this is at least equal to the $70\%$ of the total energy deposited. A cluster of $3\times 3$ crystals around this one is taken into account and the final cluster signal is computed as the sum of the nine crystal deposits. The distribution of the cluster energies are fitted with a Crystal Ball function and the widths of the Gaussian component of the Crystal ball are found to be $\sim 1.5\%$, compatible with the energy spread of the beam. The presence of the left tail is mainly due to the beam size and divergence, thus the electron may not always impact the crystal in the central region of the $3\times 3$ cluster and occasionally its shower can extend outside the nine summed crystals.
\\In August-September 2023, a pilot run was carried out at the CERN M2 muon beam line for the first time with two fully-equipped tracking station and the ECAL, this is the minimum setup to precisely measure the scattering angles of elastic interactions occurring in a thin target placed in front of the second station. A picture of the setup is shown in Fig. \ref{sett}.
The main purposes were to scale the DAQ from 1 to 2 stations, synchronize it with the ECAL DAQ and test the software and hardware alignment, with the final goal to collect enough statistics for a first measurement of the leptonic running $\Delta\alpha_{lep}$. Toward the end of the test the calorimeter was integrated with the tracker DAQ via the Serenity board, with readout at $\unita{40}{\mega\hertz}$, after a tuning of the transmitted ECAL bandwidth within the available $10$ GbE data link, together with the tracking stations.\\
The data analysis is currently ongoing and the results will be used for the technical proposal, which is planned to be submitted at the SPSC in 2024.

% Bibliography

%% [A] Recommended: using JHEP.bst file
%% \bibliographystyle{JHEP}
%% \bibliography{biblio.bib}

%% or
%% [B] Manual formatting (see below)
%% (i) We suggest to always provide author, title and journal data or doi:
%% in short all the informations that clearly identify a document.
%% (ii) please avoid comments such as "For a review'', "For some examples",
%% "and references therein" or move them in the text. In general, please leave only references in the bibliography and move all
%% accessory text in footnotes.
%% (iii) Also, please have only one work for each \bibitem.

\end{document}